\documentclass[preprint,12pt]{elsarticle}

\usepackage{amssymb}
\usepackage{amsmath}
\usepackage{caption}
\usepackage{subcaption}

\journal{} 

\begin{document}
\begin{frontmatter}
%
%
%
\title{Stress field prediction in fiber-reinforced composite materials using a deep learning approach}
%
\author[inst1]{Anindya Bhaduri}
\affiliation[inst1]{organization={Department of Civil and Systems Engineering, Johns Hopkins University}, 
            addressline={\\ 3400 N. Charles Street}, 
            city={Baltimore},
            postcode={21218}, 
            state={MD},
            country={USA}}
\author[inst1]{Ashwini Gupta}
\author[inst1]{Lori Graham-Brady}
\begin{abstract}
Computational stress analysis is an important step in the design of material systems. Finite element method (FEM) is a standard approach of performing stress analysis of complex material systems. A way to accelerate stress analysis is to replace FEM with a data-driven machine learning based stress analysis approach. In this study, we consider a fiber-reinforced matrix composite material system and we use deep learning tools to find an alternative to the FEM approach for stress field prediction. We first try to predict stress field maps for composite material systems of fixed number of fibers with varying spatial configurations. Specifically, we try to find a mapping between the spatial arrangement of the fibers in the composite material and the corresponding von Mises stress field. This is achieved by using a convolutional neural network (CNN), specifically a U-Net architecture, using true stress maps of systems with same number of fibers as training data. U-Net is a encoder-decoder network which in this study takes in the composite material image as an input and outputs the stress field image which is of the same size as the input image. We perform a robustness analysis by taking different initializations of the training samples to find the sensitivity of the prediction accuracy to the small number of training samples.
When the number of fibers in the composite material system is increased for the same volume fraction, a finer finite element mesh discretization is required to represent the geometry accurately. This leads to an increase in the computational cost. Thus, the secondary goal here is to predict the stress field for systems with larger number of fibers with varying spatial configurations using information from the true stress maps of relatively cheaper systems of smaller fiber number. 
\end{abstract}




\end{frontmatter}


\section{Introduction}
\label{sec:sample1}
Finite Element Method (FEM) \cite{bathe2006finite, reddy2010introduction} is the conventional numerical approach used for stress analysis of structures that require solving partial differential equations. 
FEM simulations can be costly when the analysis is highly nonlinear or the geometry under study is complex. Also, multi scale analyses need many computations at lower scale. A lot of work has thus been focused on replacing FEM methods by machine learning (ML) approaches that have been widely used for surrogate modeling \cite{cristianini2000introduction, williams1998prediction, bhaduri2018stochastic, bhaduri2020usefulness, bhaduri2021probabilistic} of relevant quantities of interest. 
%
%
Bock et al. \cite{bock2019review} presents a detailed overview of data mining and machine learning approaches to model process-microstructure-property-performance chain in the descriptive-predictive-prescriptive format. Applications include modeling effects of process parameters on microstructure, microstructure reconstruction \cite{li2018transfer, bhaduri2021efficient}, and capturing localized elastic strain in composites among several others. Pathan et al. \cite{pathan2019predictions} has used a gradient-boosted tree regression model to predict the homogenized properties such as macroscopic stiffness and yield strength of a unidirectional composite loaded in the transverse plane. 
Yang et al. \cite{yang2018deep} has implemented a 3D CNN architecture to predict the effective stiffness of high contrast elastic microstructures. Rao and Liu \cite{rao2020three} also utilizes a three-dimensional (3D) CNN architecture to predict the anisotropic effective properties of particle reinforced composites. Mozaffar et al. \cite{mozaffar2019deep} has used Recurrent Neural Networks (RNNs) to predict the plastic behavior of composite representative volume elements (RVEs) \cite{yvonnet2009numerically}.
Haghighat et al. \cite{haghighat2020deep} has formulated a Physics Informed Neural Networks (PINNs) framework and applied it to a linear elastostatics problem as well as a nonlinear elastoplastic problem. Liu et al. \cite{liu2021learning} introduces a deep learning based concurrent multiscale modeling approach to model the impact of polycrystalline inelastic solids. All these works mostly focus on the homogenization of properties of interest and only gives average outputs. Oftentimes, there is need to predict the variation in the local stress in order to predict local failure.\\
\indent Stress field prediction in the field of computational solid mechanics using deep learning is also an ongoing topic of research. Nie et al. \cite{nie2020stress} have implemented a deep learning approach by using two different architectures; one is the Convolutional Neural Network (FR-CNN) with a single input channel, named SCSNet, and the other is Squeeze-and-Excitation Residual network modules embedded Fully Convolutional Neural network (SE-Res-FCN) with multiple input channels, named StressNet, to predict von Mises stress field in 2D cantilevered structures. Jiang et al. \cite{jiang2021stressgan} has introduced a conditional generative adversarial network, named StressGAN, for predicting 2D von Mises stress distributions in solid structures. Sun et al. \cite{sun2020predicting} has used a modified StressNet \cite{nie2020stress} to predict the stress field for the 2D microstructure slices of segmented tomography images of a 3D fiber-reinforced polymer specimen. 
Liu et al. \cite{liu2015machine} presents machine learning approaches to predict microscale elastic strain fields in a 3D voxel-based microstructure volume element (MVE) that have potential applications in multiscale modeling and simulation of materials. 
Yang et al. \cite{yang2021deep} has developed a conditional generative adversarial neural network (cGAN) based model to predict the stress and strain field directly from the material microstructure. Sepasdar et al. \cite{sepasdar2021data} developed a two stacked generator CNN framework to predict full field damage and failure pattern prediction in composite materials.\\
\indent In this study, we are interested in predicting the local stress distribution in fiber reinforced matrix composite materials under mechanical loading using deep learning models. Specifically, there are two goals. The primary goal is to use data from a series of plane strain FEM models of a system with $N$ number of fibers to train a deep learning model that predicts the stress field in an $N$-fiber plate with arbitrary spatial distribution of $N$ fibers. A sensitivity analysis is also performed to assess the robustness of the prediction with small training sizes. The secondary goal is to predict stress field in a $M$-fiber model with varying spatial distribution of fibers using cheaper $N$-fiber model training data where $M > N$. As of now, encoder-decoder based networks \cite{nie2020stress} have proven to be efficient mappings from one image to another. Among those networks U-Net \cite{Ronneberger2015u} has captured special attention due to its ability to propagate context information from lower level layers to high level layer, making the network capable of capturing high resolution details, as well as low level features. Moreover, the skip connections have proven to be efficient to prevent the vanishing gradient problem that is associated with the training of the model. This paper implements a deep learning model with U-Net type architecture and shows its generalization capability to predict stress maps of higher number of fibers while being trained on lower number of fibers.\\
\indent This manuscript is organized as follows: Section \ref{sec:methodology} 
discusses the proposed methodology in detail. In section \ref{sec:results}, the prediction results are discussed. Section \ref{sec:conclusions} 
provides conclusions.
\section{Methodology} \label{sec:methodology}
\subsection{Problem setup}
\begin{figure}[h!]
    \centering
    \includegraphics[scale=0.35,keepaspectratio]{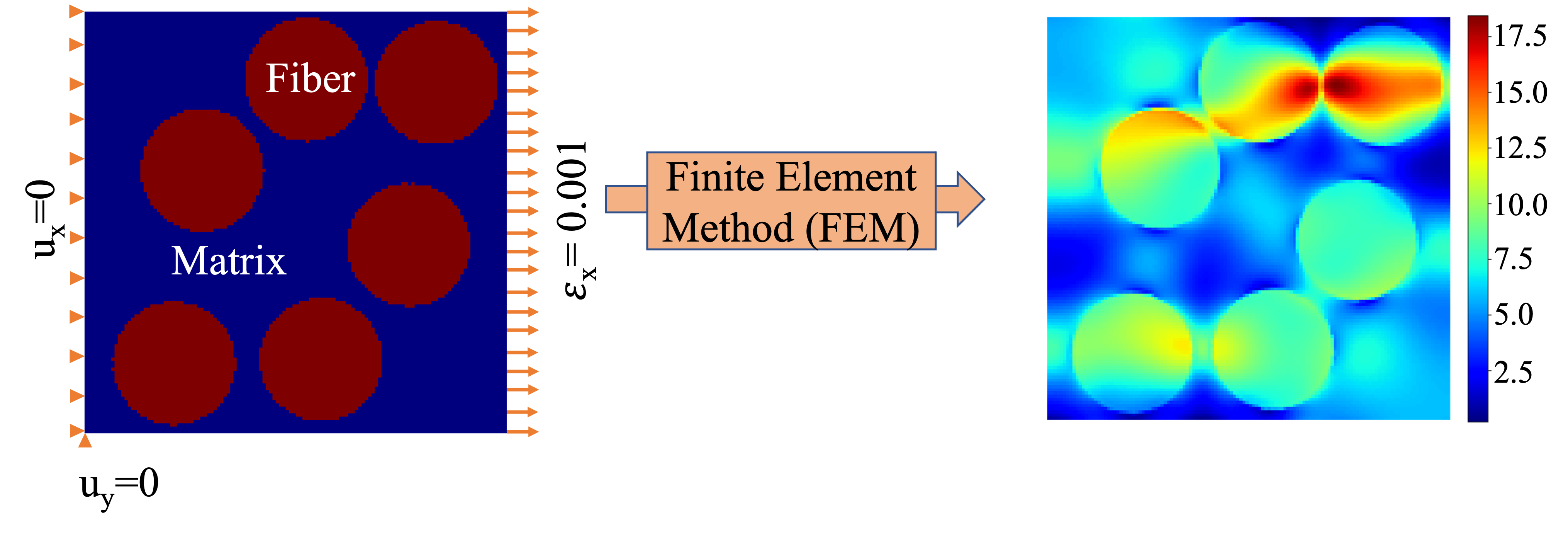}
    \caption{2D composite plate system, with horizontal applied strain (left) and the corresponding von Mises stress (right), as predicted by ABAQUS \cite{hibbett1998abaqus}.}
    \label{FEM-plate-model}
\end{figure}

A $2$-d plane strain cross-section of a fiber-reinforced composite is considered in this study as shown in figure~\ref{FEM-plate-model} where the constituent materials (fiber and matrix) are assumed linear and elastic. Under applied loading and boundary conditions, local stress and strain fields develop throughout the material. Apart from the loading and boundary conditions, this stress field depends on the mechanical properties of the matrix and the fibers, as well as the spatial distribution of the fibers, the shape of the fibers, and the volume fraction of the fiber material in the plate. 
%
%
The fibers are assumed to be circular with a fixed radius, and the volume fraction is kept constant. The fiber/matrix interface is assumed to be perfectly bonded.
The sample is subjected to a tensile strain in the horizontal direction, with traction free boundaries in the top and bottom, see figure~\ref{FEM-plate-model}. Figure~~\ref{FEM-plate-model} also shows the corresponding von Mises stress field predicted by ABAQUS \cite{hibbett1998abaqus}. The microstructural image with different spatial distribution of a particular number of fibers is the input while the output quantity of interest is the corresponding von Mises stress field under tensile load. In a sense, the FEM model provides an image-to-image mapping.
\subsection{Approach overview}
\begin{figure}[h!]
    \centering
    \includegraphics[scale=0.35,keepaspectratio]{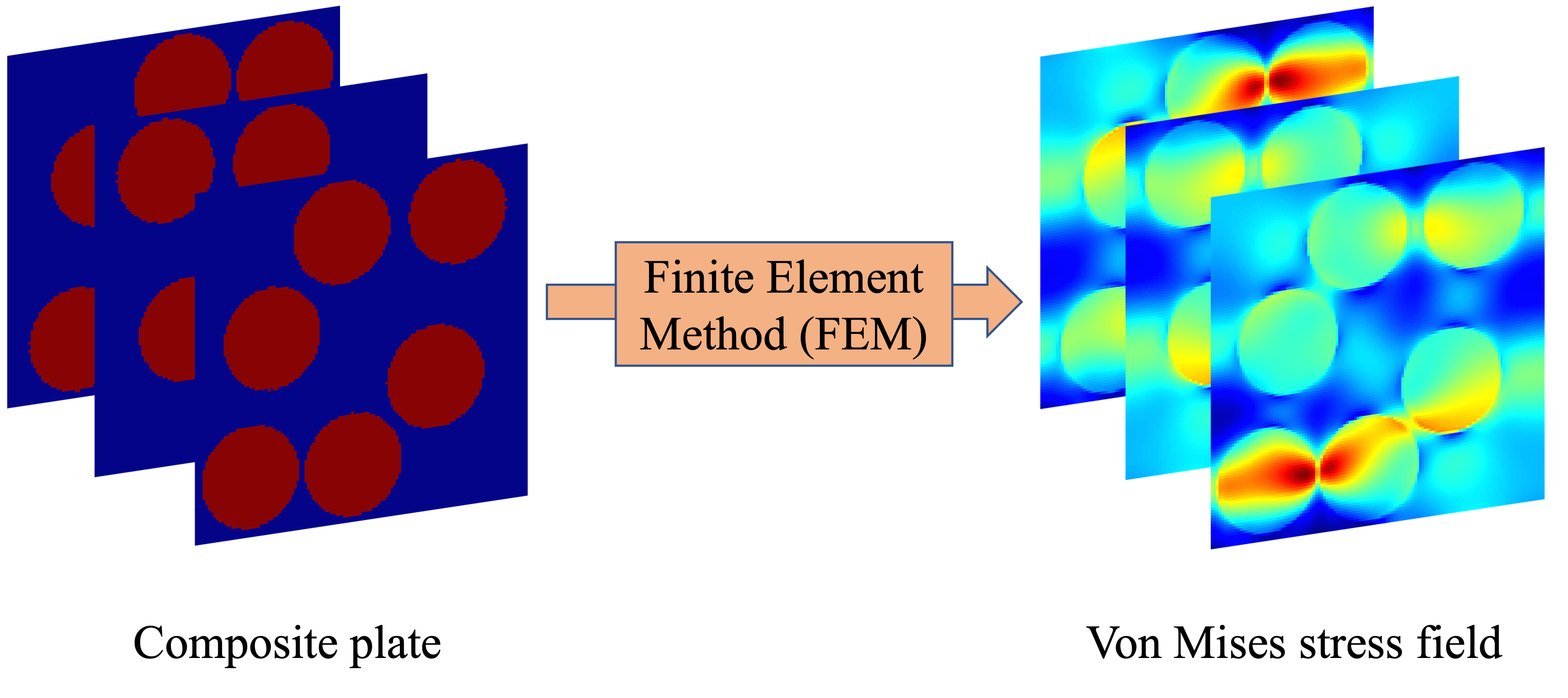}
    \caption{2D composite system composed of circular fibers embedded in a matrix along with the boundary and loading conditions (left) and corresponding von Mises stress field after a finite element method (FEM) simulation (right).}
    \label{FEM-model-run}
\end{figure}
\begin{figure}[h!]
    \centering
    \includegraphics[width=\columnwidth,keepaspectratio]{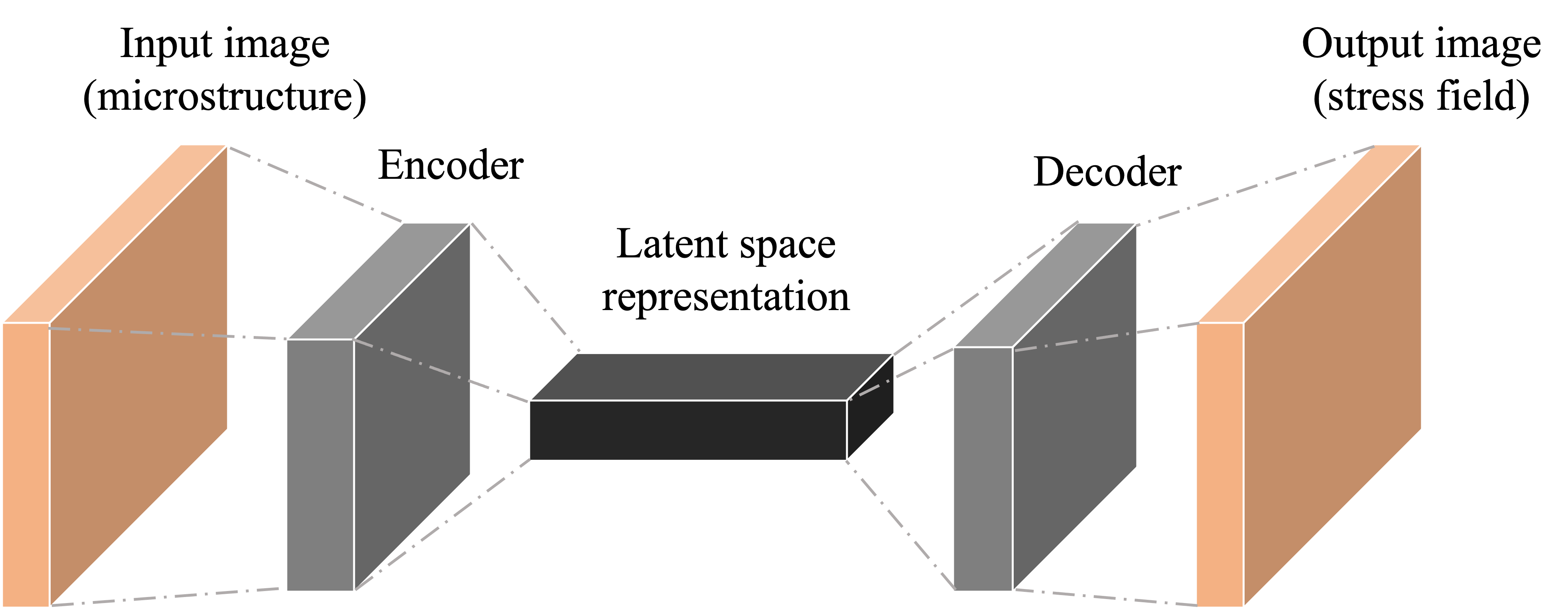}
    \caption{An encoder-decoder based network that can serve as an efficient surrogate for the FEM mapping shown in figure~\ref{FEM-model-run}}
    \label{latent}
\end{figure}
Running the FEM model for a series of microstructures provides a set of input and output image data that can be used to train a deep learning model as shown in figure~\ref{FEM-model-run}. Figure \ref{latent} shows a simple representation of the encoder-decoder network for mapping the input microstructural image to the output stress map. The input images are basically binary maps representing the location of the fibers and the matrix. The encoder-decoder network projects the input image into a lower dimensional space (called the latent space) and then projects it back to the stress field. The underlying assumption for this approach is that both the input space and the target space share the same latent space. Specifically, a U-Net \cite{Ronneberger2015u} based architecture has been used. The various weights that exist in this architecture are trained through learning based on the FEM mapping data of microstructure to stress field.
\subsection{U-Net architecture} \label{sec:unet}
The U-Net architecture was introduced initially for segmentation of medical images \cite{Ronneberger2015u}. But, over the years, it has also been proven to be efficient in capturing the latent representation for other types of images. 
\begin{figure}[b!]
    \centering
    \includegraphics[width=\columnwidth,keepaspectratio]{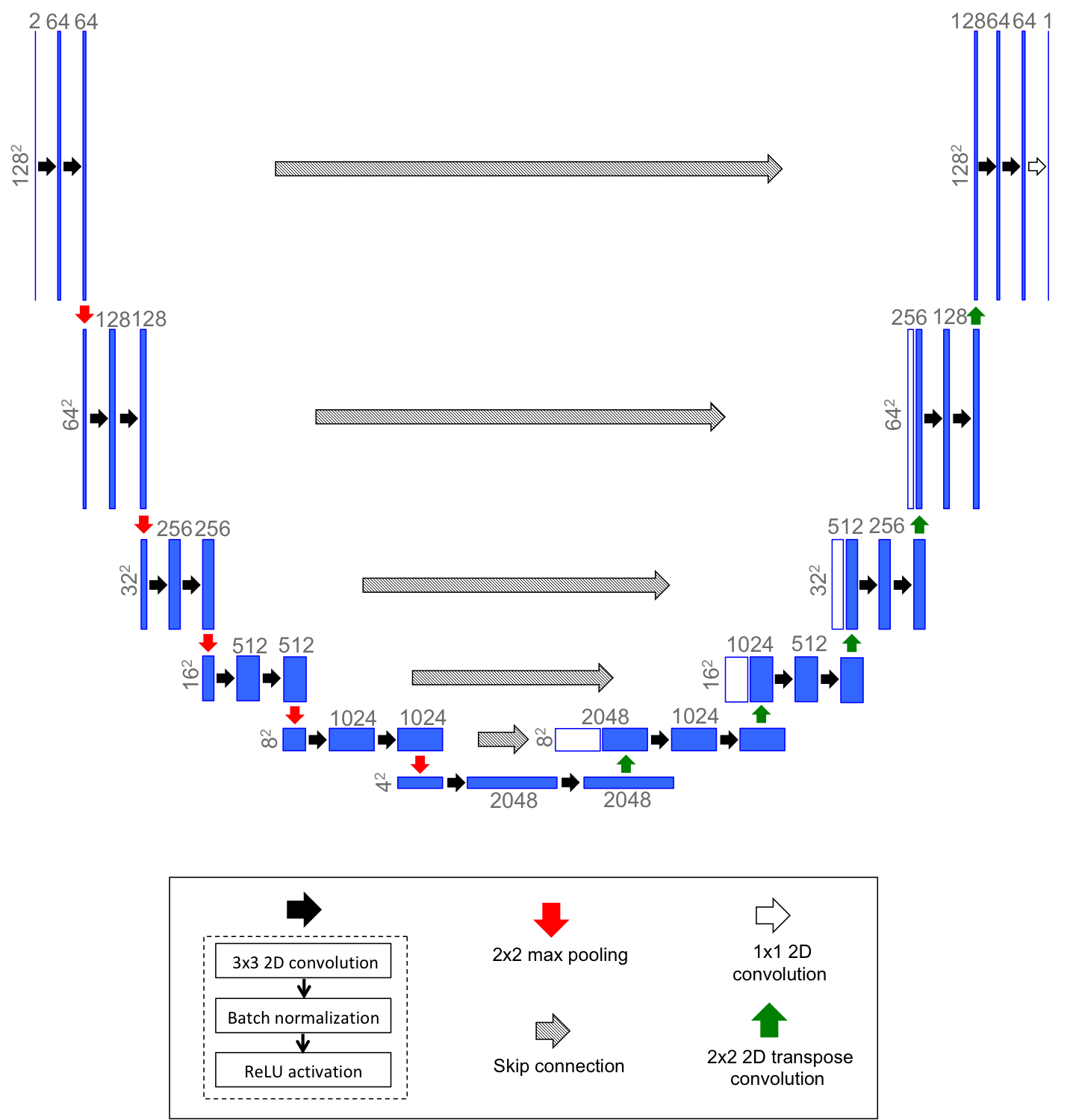}
    \caption{U-Net architecture}
    \label{unet}
\end{figure}
The standard architecture contains a series of contracting layers followed by a set of expanding layers with skip connections propagating context information from the contracting layers to the expanding layer, that enhance the resolution of the output. The U-Net architecture considered here is a slightly modified version of the original U-Net architecture \cite{Ronneberger2015u} as shown in figure \ref{unet}. The encoder part of the architecture consists of $6$ repeating blocks. Each block consists of a $2$x$2$ max pooling operation with stride 2 for downsampling, followed by application of 2 successive 3x3 2D convolutions, each followed by a batch normalization and a rectified linear unit (ReLU) operation. The first encoder block does not have the max pooling layer upfront.
The decoder part consists of 6 repeating blocks where each block (except the first and the last block) consists of two successive 3x3 2D convolutions, each followed by a batch normalization and a ReLU operation, and it is then followed by a transpose convolution operation. The first decoder block has only the transpose convolution layer and no standard convolution layers, while the last decoder block has the two successive convolution-batch norm-ReLU layers but no transpose convolution. The encoder and decoder blocks are followed by a final 1x1 convolutional layer which maps the 64-channel decoder output to a single channel. The standard U-Net requires three input layers but we use two layers in our model. Training of the weights in this architecture is done by minimizing the loss function which is taken to be the weighted mean squared error between the predicted and the true von Mises stress map from the training data.
\begin{figure}[h!]
    \centering
    \includegraphics[width=\columnwidth,keepaspectratio]{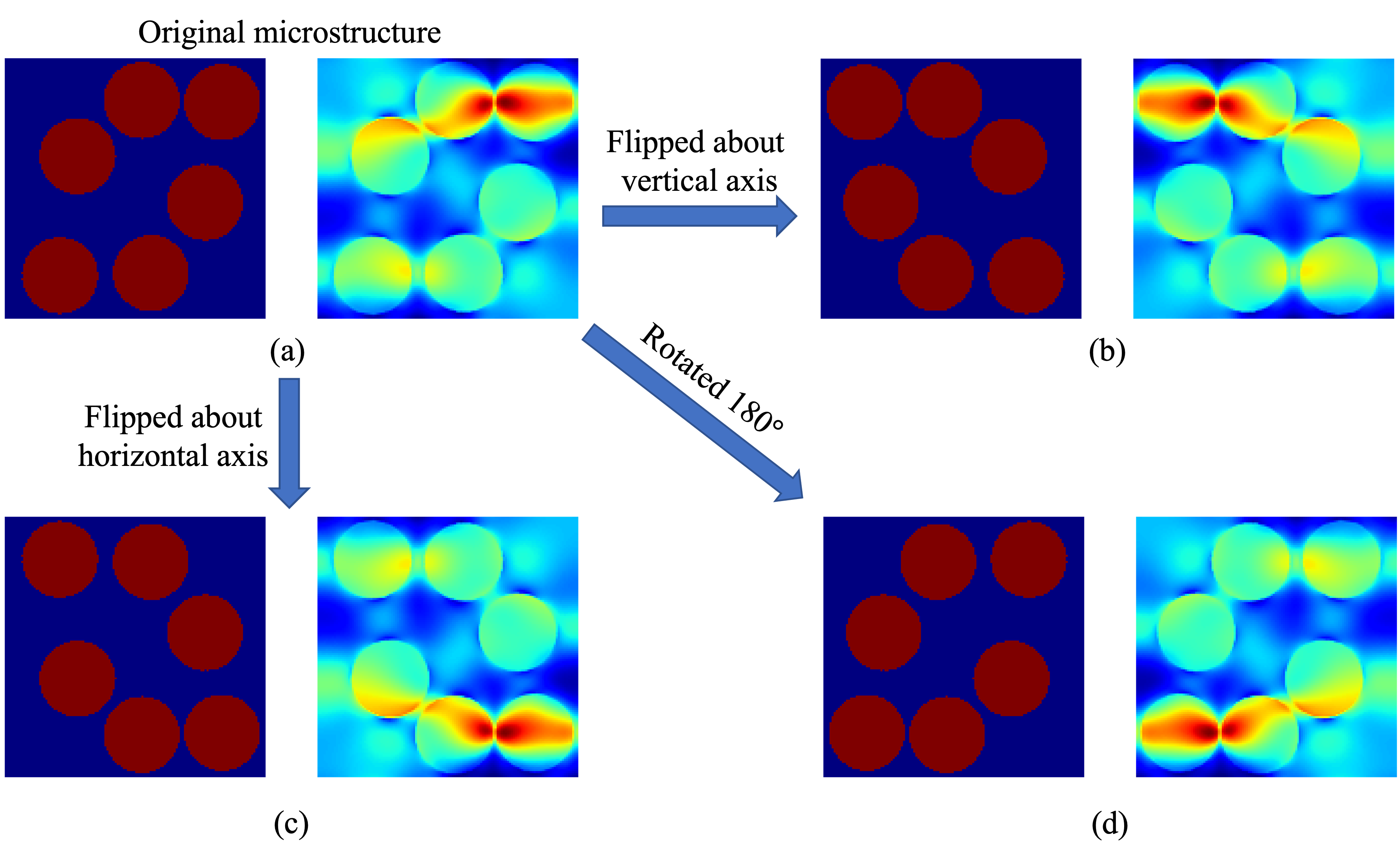}
    \caption{4-fold data augmentation by image flipping.}
    \label{data_augmentation}
\end{figure}
\section{Results} \label{sec:results}
\subsection{Stress map prediction accuracy}
\begin{figure}[b!]
    \centering
    \includegraphics[width=\columnwidth,keepaspectratio]{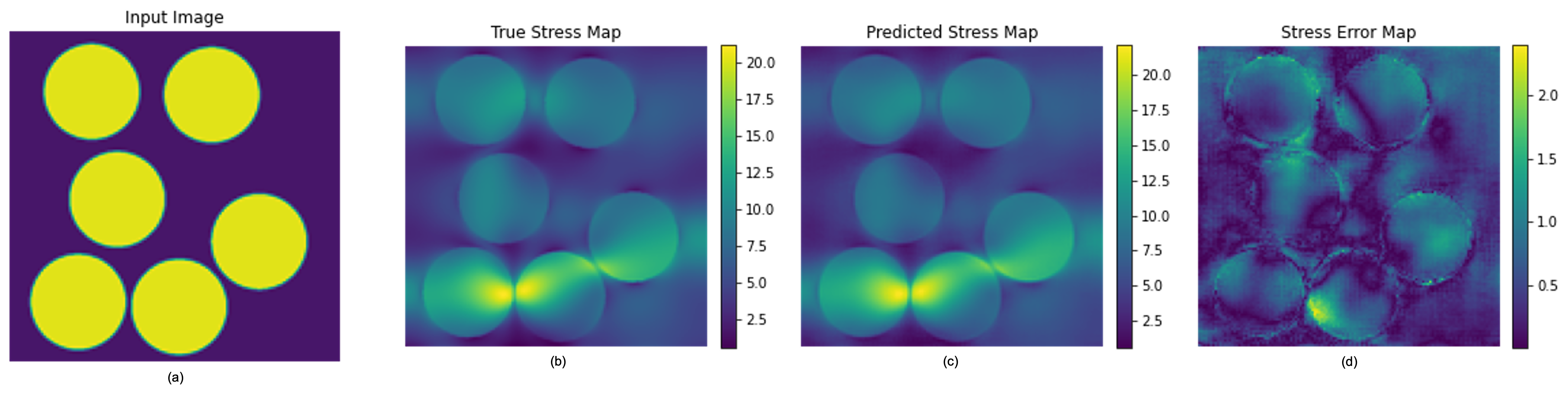}
    \caption{Von Mises stress map predicted from a U-Net architecture is based on 25 FEM analyses of a 6-fiber composite system, augmented to 100 training images.}
    \label{VonMises_6fiber}
\end{figure}
\begin{figure}[h]
    \centering
    \includegraphics[width=\columnwidth,keepaspectratio]{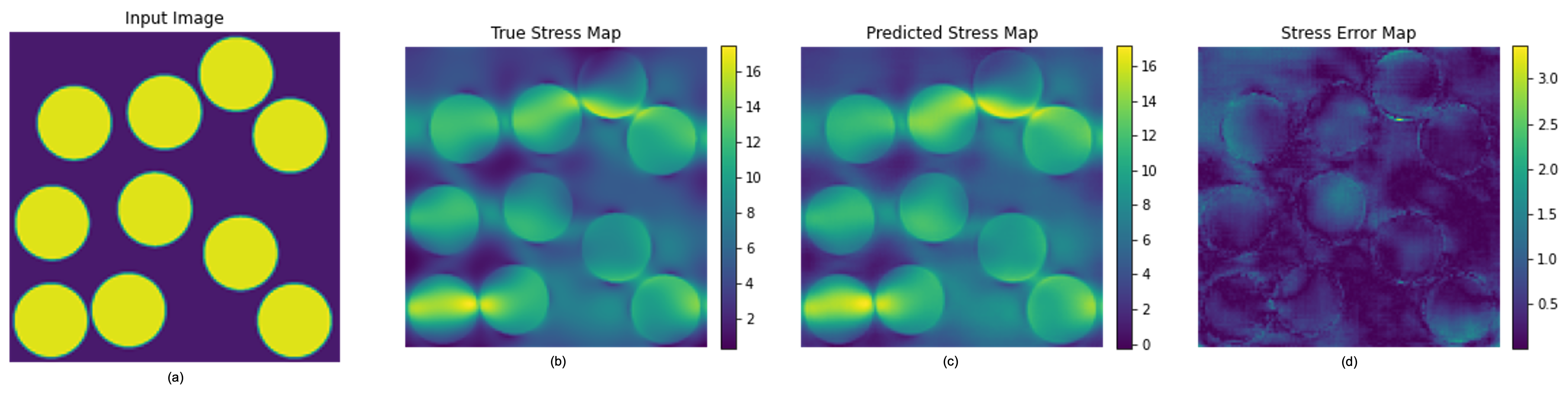}
    \caption{Von Mises stress map predicted from a U-Net architecture is based on 25 FEM analyses of a 10-fiber composite system, augmented to 100 training images.}
    \label{VonMises_10fiber}
\end{figure}
\begin{figure}[h]
    \centering
    \includegraphics[width=\columnwidth,keepaspectratio]{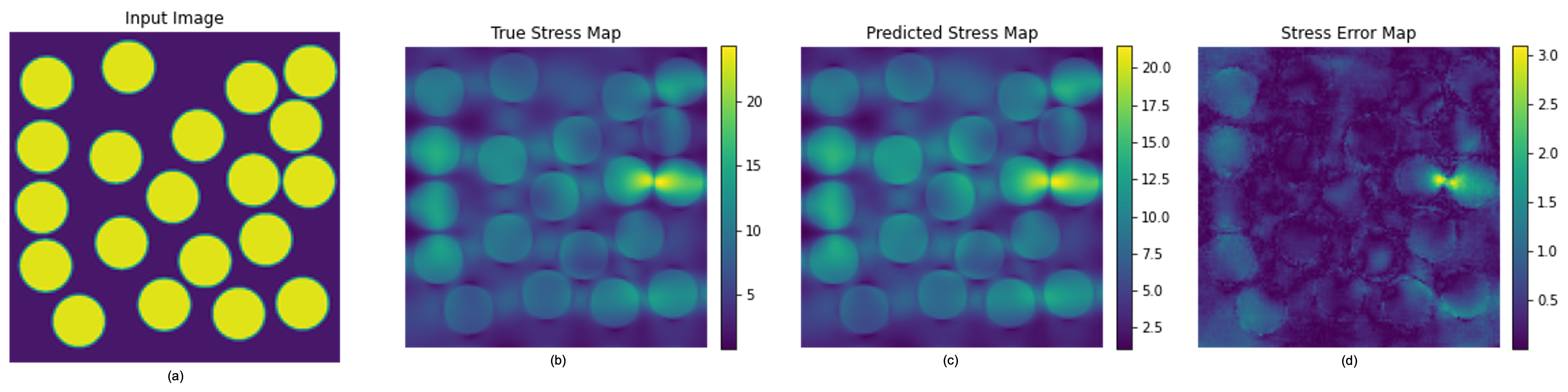}
    \caption{Von Mises stress map predicted from a U-Net architecture is based on 25 FEM analyses of a 20-fiber composite system, augmented to 100 training images.}
    \label{VonMises_20fiber}
\end{figure}
\begin{figure}[h!]
    \centering
    \includegraphics[width=\columnwidth,keepaspectratio]{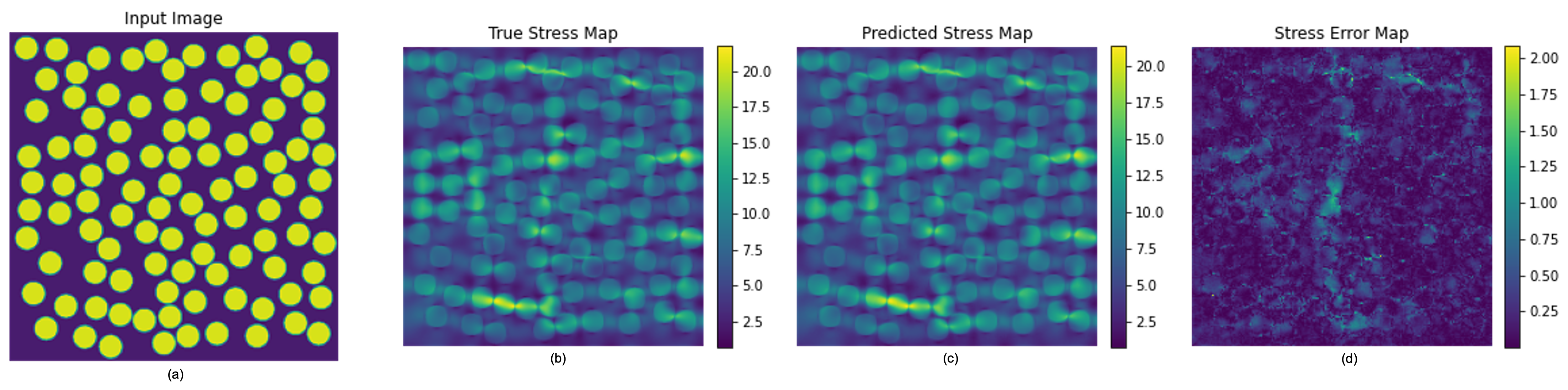}
    \caption{Von Mises stress map predicted from a U-Net architecture is based on 25 FEM analyses of a 100-fiber composite system, augmented to 100 training images.}
    \label{VonMises_100fiber}
\end{figure}

In this section, the accuracy of mapping from microstructure to the corresponding von Mises stress map is assessed by considering 6-fiber, 10-fiber, 20-fiber, and 100-fiber composite systems. Data from $25$ FEM simulations are considered for each system. Data augmentation is performed taking advantage of the physics of the problem. If the input images and the corresponding output maps are flipped appropriately, new sets of input-output data can be effectively generated. The flipping operations performed are: 1) horizontal flip, 2) vertical flip, and 3) horizontal flip followed by vertical flip. In this way, a 4-fold data augmentation has been achieved as shown in figure~\ref{data_augmentation}. After data augmentation, the training data consists of $100$ microstructural images of different spatial arrangement of fibers and their corresponding von Mises stress map for each model case. Figure \ref{VonMises_6fiber}a shows the input microstructural image for the 6-fiber composite system. Figure \ref{VonMises_6fiber}b and \ref{VonMises_6fiber}c shows the true (FEM simulated) and predicted (U-Net learned) von Mises stress maps. Figure \ref{VonMises_6fiber}d shows the corresponding stress error map which indicates that the prediction error is relatively small. Figures \ref{VonMises_10fiber}, \ref{VonMises_20fiber} and \ref{VonMises_100fiber} show similar plots for the 10-fiber, 20-fiber, and 100-fiber composite systems respectively which also indicate good stress map prediction performance of the U-Net. The prediction error is quantified in the following section.
\subsection{Effect of training size on stress map accuracy}
The quality of the predicted stress maps varies with the training data size. 
In order to assess this, training is performed 20 times, each with different random seed initializations that led to 20 different training datasets. 4 different accuracy metrics are considered, namely the weighted mean squared error (weighted mse), the mean maximum error, the median maximum error and a normalized root mean squared error (RMSE/range). If the height and width of the images are denoted by $H$ and $W$, then the size of the images are given by $S = H \times W$. Let $N_{test}$ denote the number of test images considered. The mean weighted MSE is calculated by using the true stress values at each pixel in the test image as weights to estimate a weighted mean squared error for each test image and then taking the mean over all the test images. It is then defined as:
\begin{equation}
    \text{Mean weighted MSE} = \frac{1}{N_{\text{test}}}\sum_{j=1}^{N_{\text{test}}} \left[ \frac{\sum_{i=1}^{S} y_t^i (y_p^i - y_t^i)^2}{\sum_{i=1}^{S} y_t^i} \right]^j
\end{equation}
where $y_t^i$ is the true stress value at pixel $i$ of a test image and $y_p^i$ is the corresponding predicted stress value at the same pixel $i$.\\
\indent The median maximum error is calculated by measuring the difference in the maximum true and predicted stress values in each test image and taking the median of that quantity over all the test images. It is thus defined as:
\begin{equation}
    \text{Median maximum error} = \text{median}_j \left[ |y_p^{\text{max}} - y_t^{\text{max}}| \right]^j, \ \ \  j = 1, \dots, N_{\text{test}}
\end{equation}
where $y_t^{\text{max}}$ is the true maximum stress value of a test image and $y_p^{\text{max}}$ is the corresponding predicted maximum stress value.\\
\indent The mean maximum error is calculated by measuring the difference in the maximum true and predicted stress values in each test image and taking the mean of that quantity over all the test images. It is thus defined as:
\begin{equation}
    \text{Mean maximum error} = \frac{1}{N_\text{{test}}}\sum_{j=1}^{N_\text{{test}}}\left[|y_p^{\text{max}} - y_t^{\text{max}}|\right]^j
\end{equation}
\indent The normalized RMSE is calculated by measuring the RMSE over all the test images and dividing it by the true range of the stress values over all test images. It is defined as:
\begin{equation}
    \text{Normalized RMSE} = \frac{1}{\text{R}}\sqrt{\frac{1}{N_\text{{test}}}\sum_{j=1}^{N_{\text{{test}}}} \left[ \frac{\sum_{i=1}^{S} (y_p^i - y_t^i)^2}{S} \right]^j}
\end{equation}
where R = $\text{max}_{j} \left[y_t^{\text{max}}\right]^j - \text{min}_{j} \left[y_t^{\text{min}}\right]^j (j = 1, \dots, N_{\text{test}}$) denotes the true range.\\
\begin{figure}[t!]
    \centering
    \includegraphics[width=\columnwidth,keepaspectratio]{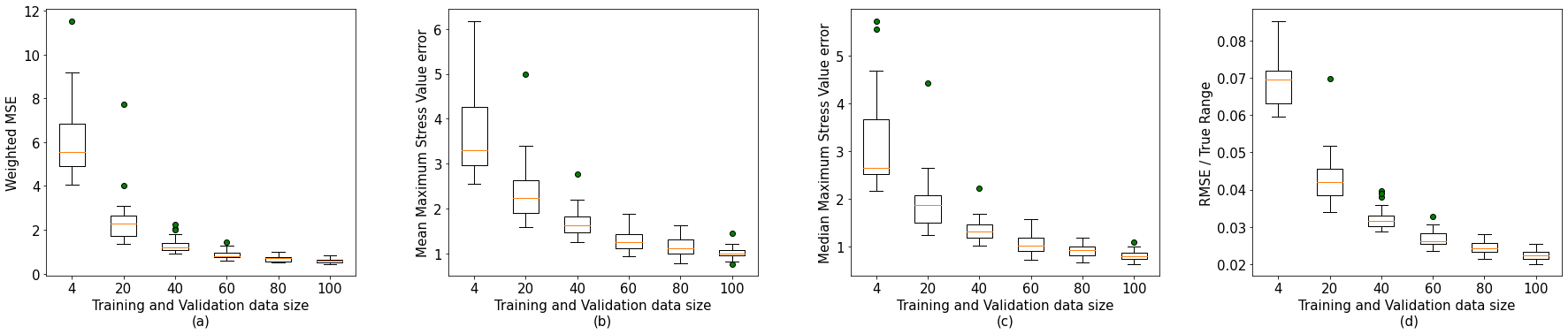}
    \caption{Accuracy metric convergence for 6-fiber composite system.}
    \label{ErrConvPlot_6fib}
\end{figure}
\begin{figure}[t!]
    \centering
    \includegraphics[width=\columnwidth,keepaspectratio]{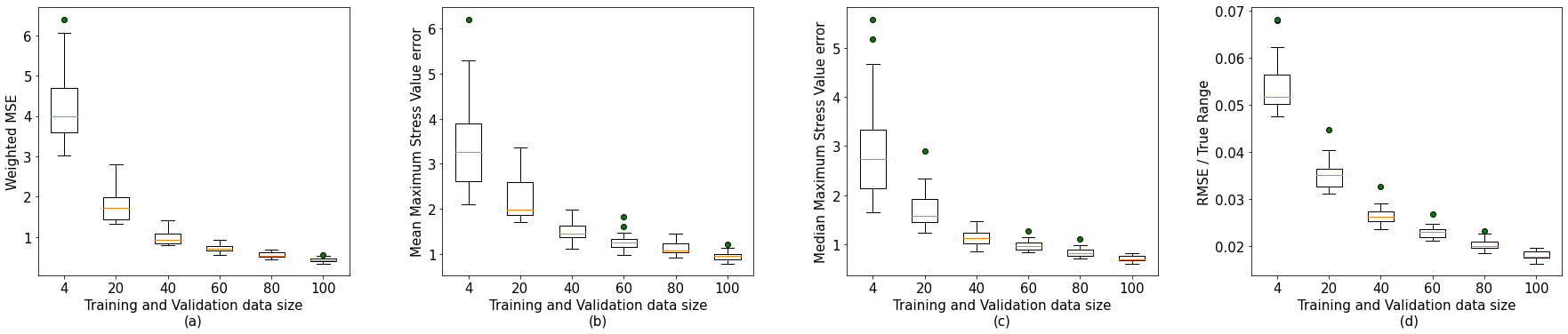}
    \caption{Accuracy metric convergence for 10-fiber composite system.}
    \label{ErrConvPlot_10fib}
\end{figure}
\begin{figure}[t!]
    \centering
    \includegraphics[width=\columnwidth,keepaspectratio]{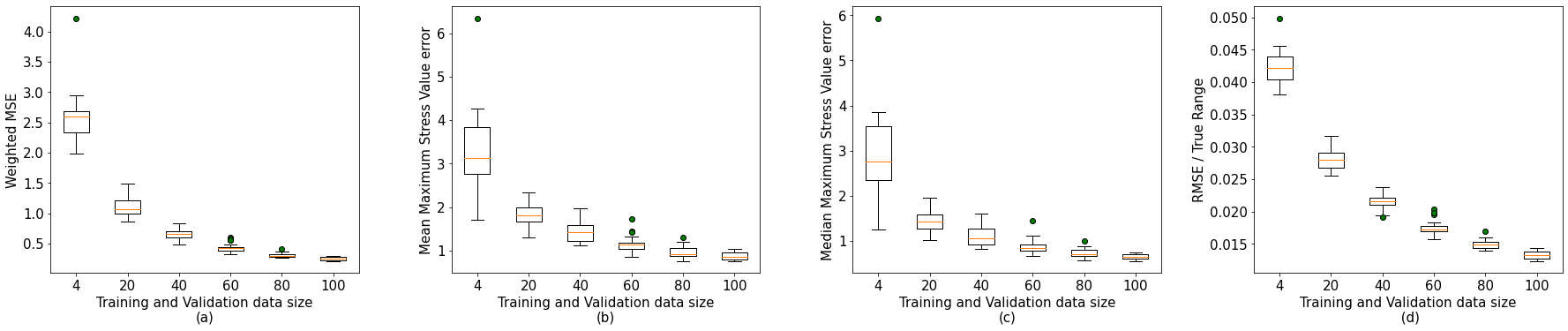}
    \caption{Accuracy metric convergence for 20-fiber composite system.}
    \label{ErrConvPlot_20fib}
\end{figure}
\begin{figure}[h!]
    \centering
    \includegraphics[width=\columnwidth,keepaspectratio]{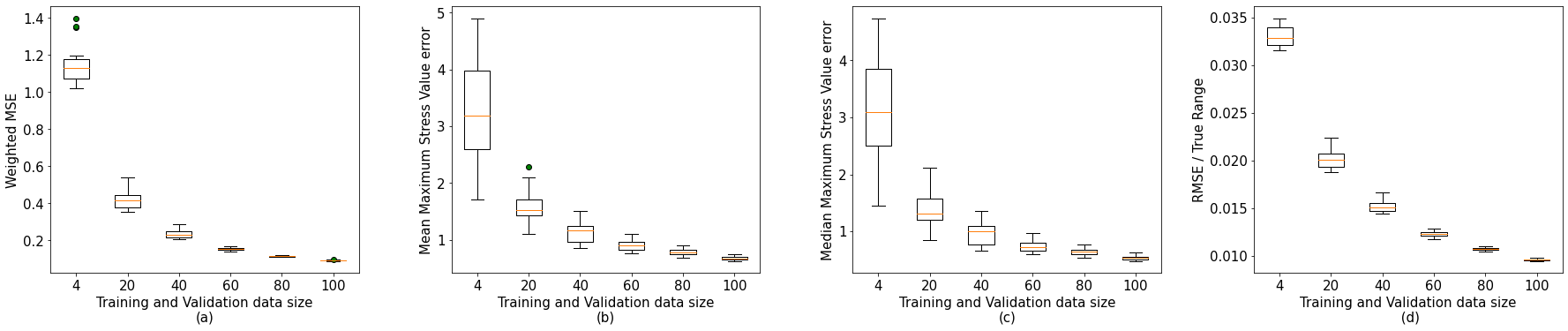}
    \caption{Accuracy metric convergence for 100-fiber composite system.}
    \label{ErrConvPlot_100fib}
\end{figure}
As expected, with increase in training data size, the mean accuracy increases and the variance of the accuracy decreases. Figures \ref{ErrConvPlot_6fib}, \ref{ErrConvPlot_10fib}, \ref{ErrConvPlot_20fib}, and \ref{ErrConvPlot_100fib} show the error bar plots of the above mentioned error metrics for 6-fiber, 10-fiber, 20-fiber and 100-fiber systems. It is also observed that even though the stress field magnitudes are similar across the different composite systems, the overall error across all metrics decreases with increase in the number of fibers from $6$ to $100$. This is attributed to the fact that with increase in number of fibers in the composite, the higher stress values are localized over a smaller region.
\begin{figure}[b!]
    \centering
    \includegraphics[scale=0.2,keepaspectratio]{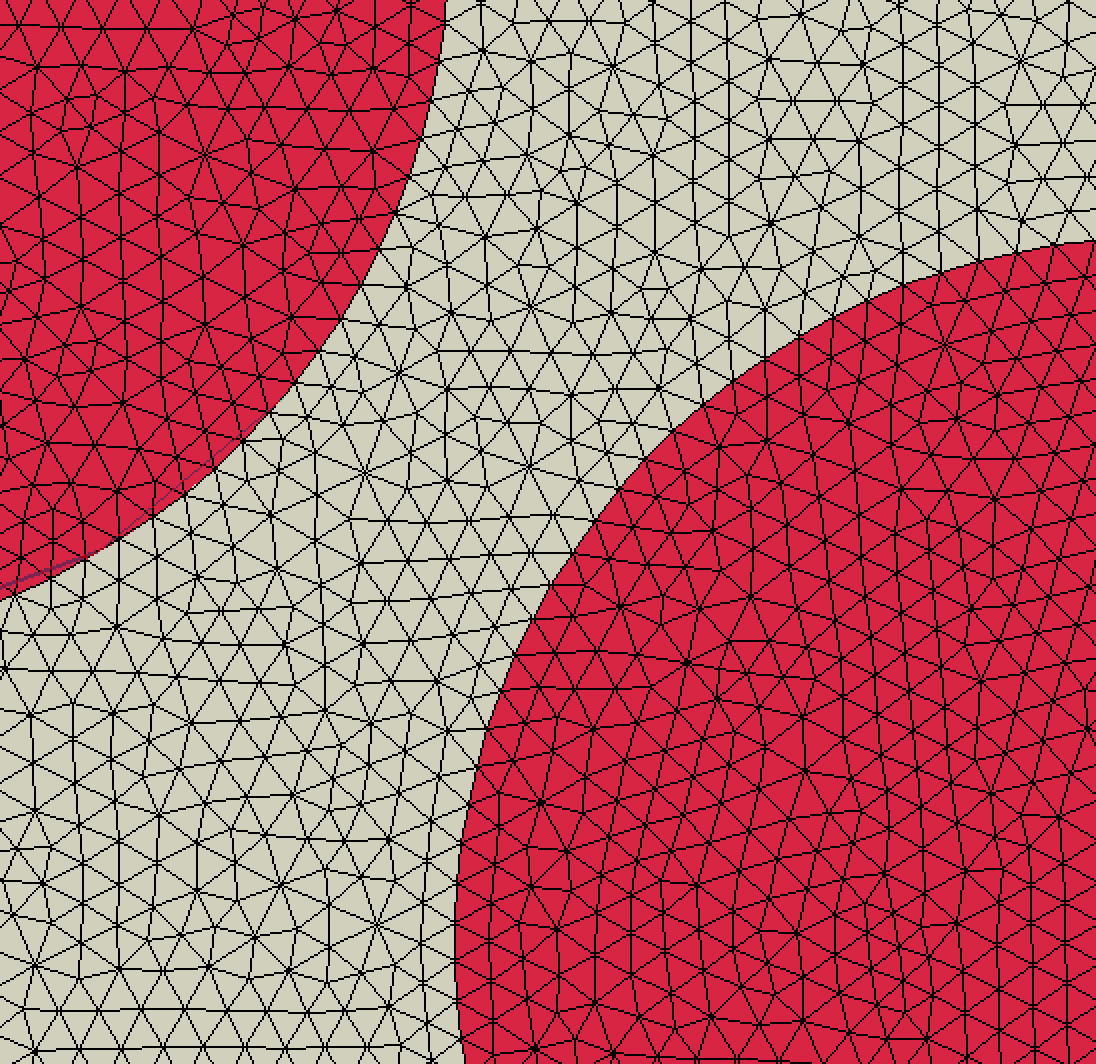}
    \caption{FEM mesh resolution}
    \label{element-resolution}
\end{figure}
\subsection{Deep transfer learning}
Transfer learning is an efficient approach where previously learned deep learning model weights are used as initial weights for retraining the same model with a smaller amount of new data, which can help to achieve faster convergence for the new dataset. This concept is evaluated for the composite material problem under study, specifically using a U-Net architecture trained on a composite system with a certain number of fibers to reduce the training effort for another composite system with a higher number of fibers. In the $2$-d fiber-reinforced composite, the finite elements must be small enough to resolve the region around the fiber/matrix interface as shown in figure~\ref{element-resolution}. For a fixed volume fraction, if the number of fibers is increased, the fibers and interfiber spacings reduce in size and the number of finite elements must therefore increase to have a good quality mesh. A sample with more fibers is therefore more expensive to solve. This motivates the use of a transfer learning approach to predict stress maps in expensive composite systems with a higher number of fibers.\\
\begin{figure}[h!]
     \centering
     \begin{subfigure}[b]{1.0\textwidth}
         \centering
         \includegraphics[width=\textwidth]{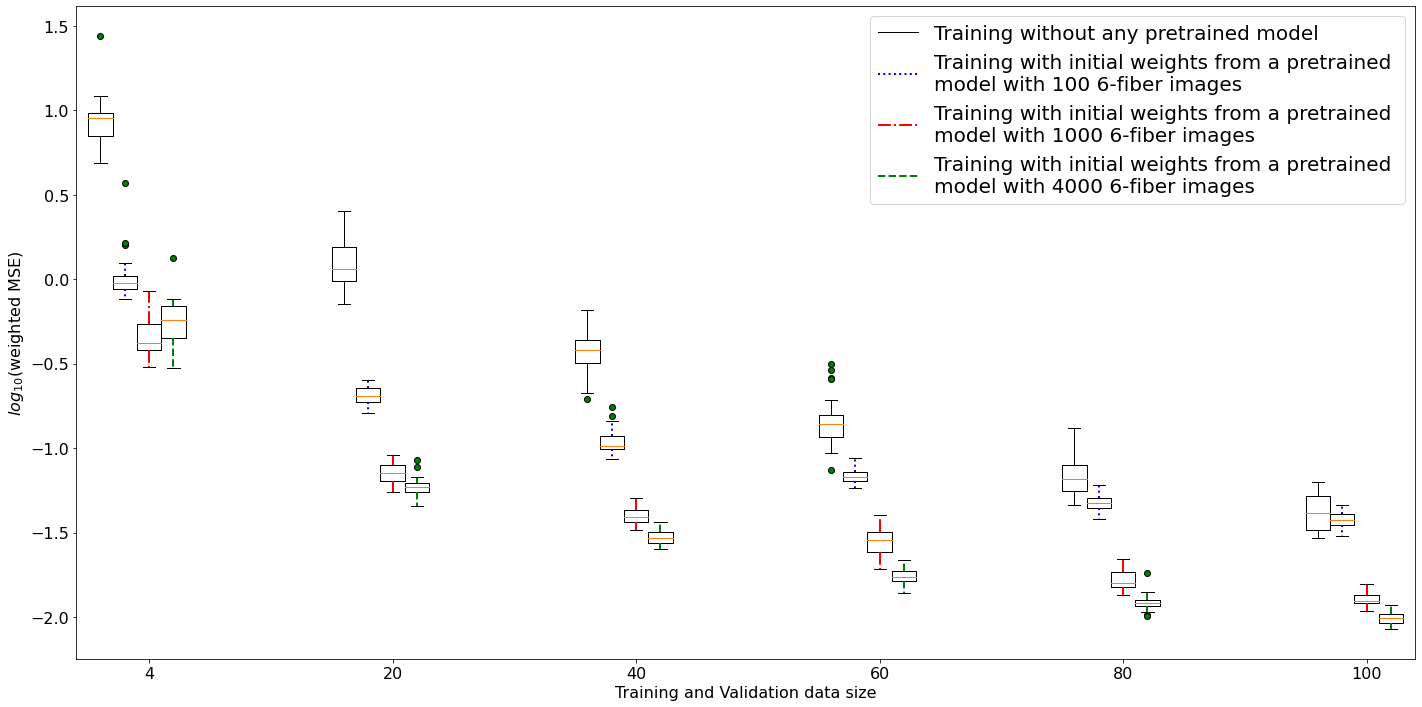}
         \caption{20-fiber composite system}
         \label{fig:20fib_transferlearning}
     \end{subfigure}
     \hfill
     \begin{subfigure}[b]{1.0\textwidth}
         \centering
         \includegraphics[width=\textwidth]{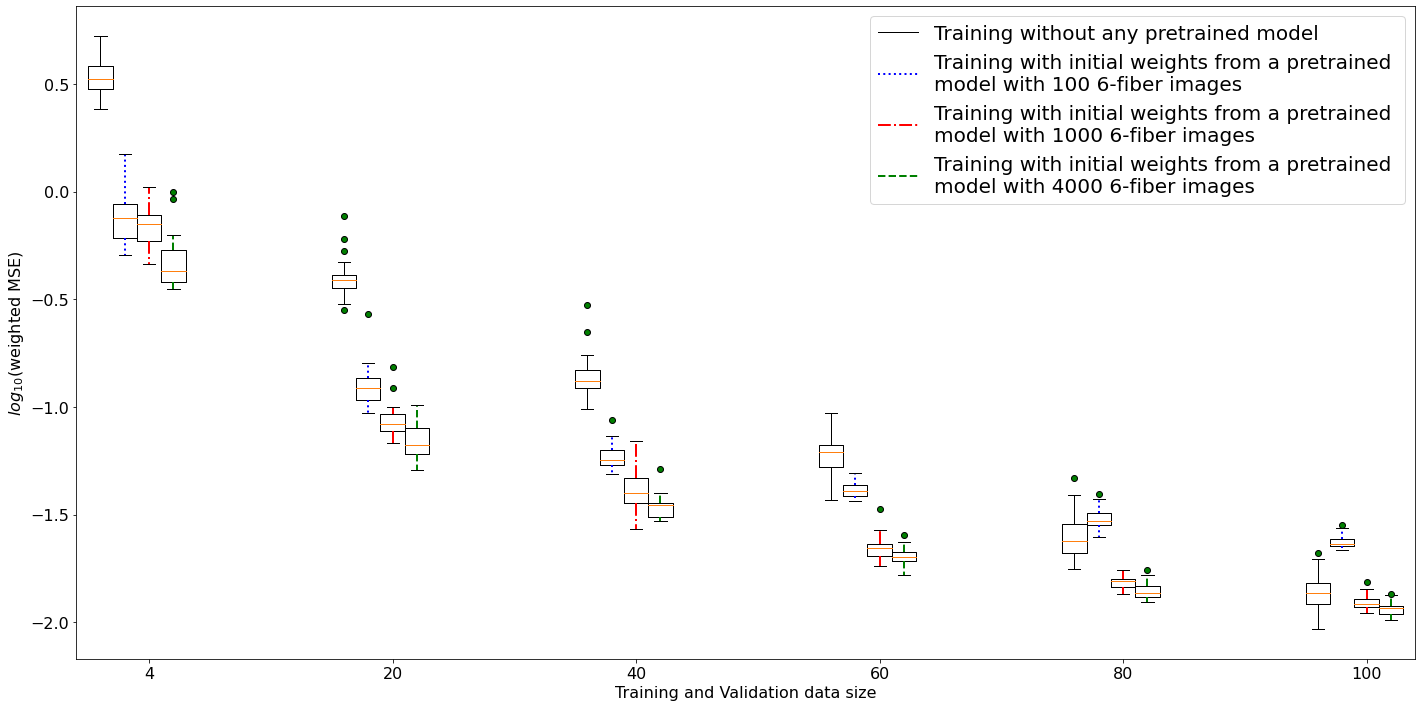}
         \caption{50-fiber composite system}
         \label{fig:50fib_transferlearning}
     \end{subfigure}
        \caption{Prediction on 20-fiber and 50-fiber composite systems based on transfer learning from 6-fiber composite system}
        \label{fig:transfer_learning}
\end{figure}
\indent In particular, information from a 6-fiber system trained model is used to predict the von Mises stress map of systems with 20 and 50 fibers. Figure \ref{fig:20fib_transferlearning} shows the transfer learning results for the 20-fiber composite systems. If the U-Net is trained with only 20-fiber composite system data from scratch, then the prediction accuracy is worse than the case where the training is performed using a pretrained U-Net model with 6-fiber system data. For most cases, the results improved when the U-Net model was pretrained on more 6-fiber system data, although the improvement decreases as more data is used. Figure \ref{fig:50fib_transferlearning} shows the transfer learning results for the 50-fiber composite systems. It is seen in this case that transfer learning helps in achieving better accuracy for the smaller training data size, but with data sizes of 80 and 100, there is little or no advantage with pretraining information. In fact, when data size of 100 is used for the 50-fiber system, the results based on the pretrained U-Net model with 100 6-fiber system data lead to a prediction accuracy that is worse than the results from the U-Net model trained from scratch. The ineffectiveness of the pretrained information can be attributed to the fact that the stress distribution in the 50-fiber composites is quite different from that of the 6-fiber composites.
\section{Conclusions} \label{sec:conclusions}
In this study, a U-Net architecture has been used to accurately predict the von Mises stress field for 6-fiber, 10-fiber, 20-fiber, and 100-fiber composite plates under uniaxial tension with an arbitrary spatial arrangement of fibers. A weighted mean square loss function has been used to predict high stress regions accurately. A sensitivity analysis was performed to evaluate the accuracy of prediction with different size of the training data, confirming that results improve with increased data. A transfer learning approach was used to predict the stress distribution in 20-fiber and 50-fiber systems using a U-Net model pretrained with 6-fiber system data. In almost all the cases, the pretrained network gave superior accuracy. This serves as an example of the applicability of deep learning architectures in stress field prediction. \\
\indent A direct extension of this presented work is to obtain stress map predictions of all the stress components, instead of just the von Mises stress, as a function of an arbitrary strain vector. This can provide a path towards an efficient ML driven multi scale model.
Another future direction is to relax some of the simplifying assumptions incorporated in the composite systems considered in this study. For example, fiber/matrix interfacial debonding as well as damage in the constituent phases can be considered which then becomes a history-dependent problem and is much more complicated to deal with. The challenge there will be to accurately predict the evolution of damage as well as the stress field with time under a given load.
\section*{Acknowledgements}
Research was sponsored by the Army Research Laboratory and was accomplished under Cooperative Agreement Number W911NF-12-2-0023 and W911NF-12-2-0022. The views and conclusions contained in this document are those of the authors and should not be interpreted as representing the official policies, either expressed or implied, of the Army Research Laboratory or the U.S. Government. The U.S. Government is authorized to reproduce and distribute reprints for Government purposes notwithstanding any copyright notation herein.


\bibliographystyle{elsarticle-num} 
\bibliography{cas-refs}





\end{document}